# Crater-like landform in Bayuda desert (a processing of satellite images)

**Amelia Carolina Sparavigna**
Dipartimento di Fisica, Politecnico di Torino
C.so Duca degli Abruzzi 24, Torino, Italy

**Abstract**
The paper is proposing a survey of a region in Sudan, the Bayuda desert, using the satellite images as obtained from Google Maps. The images reveal the ring granitic structure of the region enclosed by a bend of river Nile. To enhance the features of the landform, images are processed with a method based on fractional calculus, able to increase the rendering of edges without deteriorating the overall quality of images. Besides the ringed structure of the region, a huge crater-like structure with a diameter of 10 km is evidenced.

**Key-words**: Satellite maps, Crater landform, Image processing, Fractional calculus.

The paper is proposing a survey of the Bayuda desert, an arid region in Sudan outlined by a bend of the river Nile, examining the satellite Google Maps. This kind of activity was stimulated by the discovery of a small impact crater in the Egyptian desert, the Kamil crater, which was in fact observed and located in 2008 using satellite imagery in Google Earth by Vincenzo de Michele [1,2]. Of course, craters created by the impacts of celestial bodies on Earth or by volcanic or other activities, have a clear shape in the satellite imagery when images are showing locations in deserts.
The desert of Bayuda is placed in the huge bend, from the fourth to the sixth cataract, of the river Nile (see Fig.1). It is characterised from the basaltic rocks of ancient volcanoes. Photographic and geological studies of Bayuda volcanic field and of Jebel Abu Nahl, at its approximate center, have revealed a Precambrian origin of the ground [3,4]. In Ref. [3], the authors show that a cluster of well-preserved volcanoes are present in the northern part of Bayuda Desert. These volcanoes are creating a more or less continuous field, with isolated centres of eruption. According to the authors, the distribution of volcanoes was partly controlled by large granitic ring-intrusions. These intrusions belong to the Younger Granite association of Late Precambrian or Lower Palaeozoic age and represent one of the volcanic-intrusive episodes widespread in northern Africa. In Ref.[4], photographic and geologic studies of Jebel Abu Nahl proposed this location as a Late Precambrian ring-complex, formed as a result of near-surface volcanic activities. The authors have observed the subsidence of surface or near-surface blocks along arcuate ring-faults. This part of Sudan is then attractive for the geologic studies, and considered as a part of the Nile Basin of "supreme importance" [5], because of its very long geological history. As recently proposed, it is a part of the Gondwanaland and hosted the position of the North Pole, as determined by studies on the magnetic directionality of some of the rocks [6].
References [3] and [4] displayed the ring granitic structure of the Bayuda desert. A well-known example of such structure is the Jebel Uweinat, a region at the Egyptian-Libyan-Sudanese border. Its western part consists of granite, arranged in a ring shape of some 25 km diameter. Let us then observe the Bayuda region on Google maps. As we can see from Fig.1, it seems that arched textures are present in the landform. To increase the visibility of these textures we used an image processing tool (AstroFracTool), based on fractional calculus, previously used for astronomical images to enhance the edges without loosing the whole image texture [7]. A further slight adjustment of contrast and brightness with Gimp of the map obtained after AstroFracTool application gives the result shown in Fig.2. Note the embossment effect, due to the use of AstroFracTool, of the paths of rivers.
In the image, two regions of the Bayuda desert are outlined, because of their bent ringed texture. The region on the west side of Nile shows inside a well-defined ring with a diameter of approximately 10 km, the region on the east side of Nile has a non-perfectly round shape, but it is huge (approx. 40 km diameter). Let us investigate the first region with more detail. Images in Fig.3, (a) and (c), are then referring to the region on the west side of river. In the right panels of the Figure, the images, (b) and (d), obtained after enhancement with AstroFracTool and Gimp, acquire an embossment effect. A crater-like landform, with its almost perfect circular shape, is clearly visible. To the author's knowledge this structure is not discussed in literature. This crater is approximately 40 km west of Berber town.

The following Fig.4 is showing in detail the region east side of Nile.

We could compare the images we obtained with those of Kebir crater field in Egypt or Arkenu craters in Libya. Arkenu are a pair of eroded craters, with diameters of 10 km and 6.8 km, believed to have been formed simultaneously as a double impact event and exposed by erosion at the surface [8]. Other scientists are debating the origin of Arkenu [9]: they carried out a preliminary geological and structural survey which is not supporting the idea that Arkenu were impact craters. The same authors proposed a non-impact origin for the circular structures in Gilf Kebir area (S-W Egypt), previously identified as an impact crater field [10]. In [9], the authors prefer to define the Arkenu craters as "pseudo-shatter cones", where shatter cones are the rare landforms that are created in the bedrock beneath impact craters or underground nuclear explosions. The authors are guessing that these cones are the result of intrusion of a paired, nearly cylindrical sub- volcanic stocks accompanied by degassing and followed by further structural adjustments.

For the author of the present paper, a further discussion on the origin of the crater-like landform in Bayuda desert, impact crater or volcanic, is impossible. The aim of this paper was that of showing the possibility to enhance landforms in satellite images with processing methods based on fractional calculus, to reveal known and unknown features of the ground. Moreover, the paper is evidencing a huge crater-like structure in the observed region of Bayuda desert.

**Figure captions**

. Fig.1 The bend of the river Nile from the fourth to the sixth cataract. This desert region of Sudan, the Bayuda desert, is characterised from its basaltic rocks of ancient volcanoes. The Gebel Abu Nahl is dominating the bend of the river. The image is obtained from Google Maps.

Fig.2 The same region as in Fig.1, with an enhancement of details obtained with fractional image processing. The marked regions have a bent terrain. The region on the west side of Nile shows a crater-like structure with a diameter of approximately 10 km. The other marked region on the east side of Nile has a non-perfect round shape, but it is huge (approx. 40 km diameter). The image was obtained processing Google Map images.

Fig.3 The image shows in detail the region on the west side of river Nile (see Fig.2). The crater-like landform, with its coordinates marked, in the upper left panel (a) has a diameter of approximately 10 km. This structure is 40 km west of Berber town. The panel (a) shows the image as obtained from Google Maps. On the right (b), the image after enhancement with AstroFracTool and Gimp. Note the embossment effect due to the use of AstroFracTool. It seems that at contact with this crater-like structure, there is another ringed structure, with a subtle bright edge, but this form could be a not genuine one. The two lower panels are showing the crater enlarged: on the left (c) the original image and on the right (d) the processed one.

Fig.4 This is a detail of the region on the east side of Nile which is not a perfectly round (approx. 40 km diameter). This image was obtained processing a Google Map

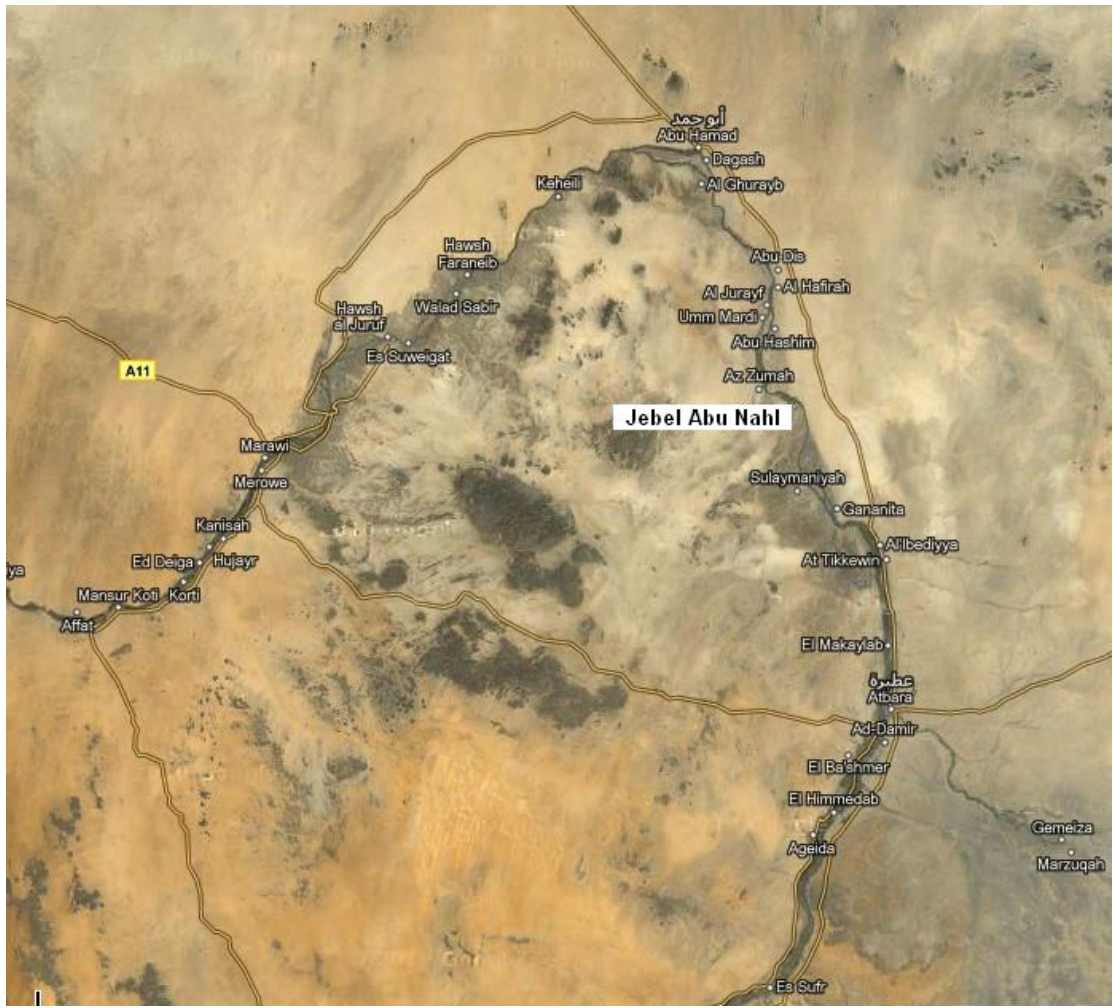

. Fig.1 The bend of the river Nile from the fourth to the sixth cataract. This desert region of Sudan, the Bayuda desert, is characterised from its basaltic rocks of ancient volcanoes. The Gebel Abu Nahl is dominating the bend of the river. The image is obtained from Google Maps.

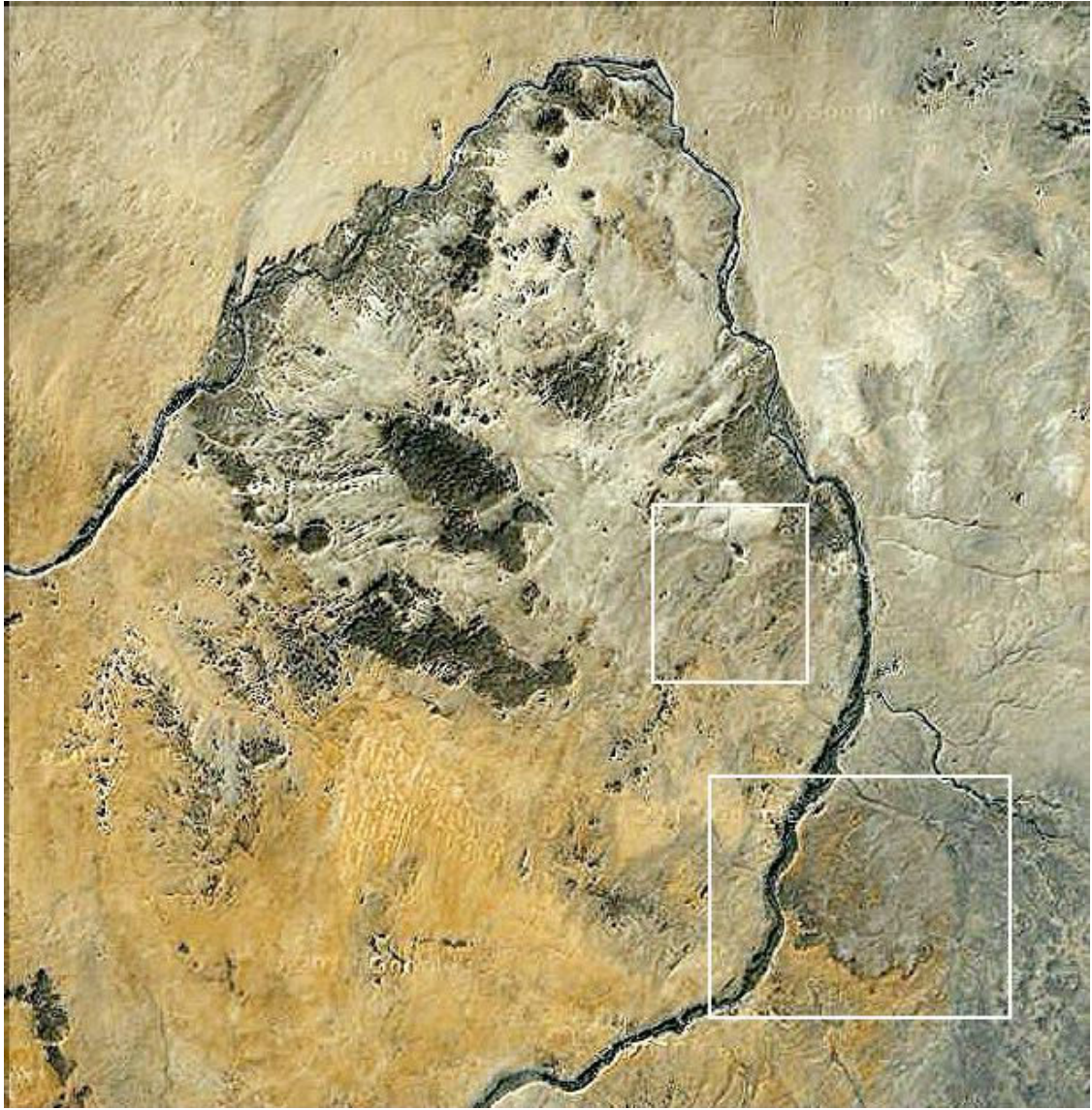

Fig.2 The same region as in Fig.1, with an enhancement of details obtained with fractional image processing. The marked regions have a bent terrain. The region on the west side of Nile shows a crater-like structure with a diameter of approximately 10 km. The other marked region on the east side of Nile has a non-perfect round shape, but it is huge (approx. 40 km diameter). The image was obtained processing Google Map images.

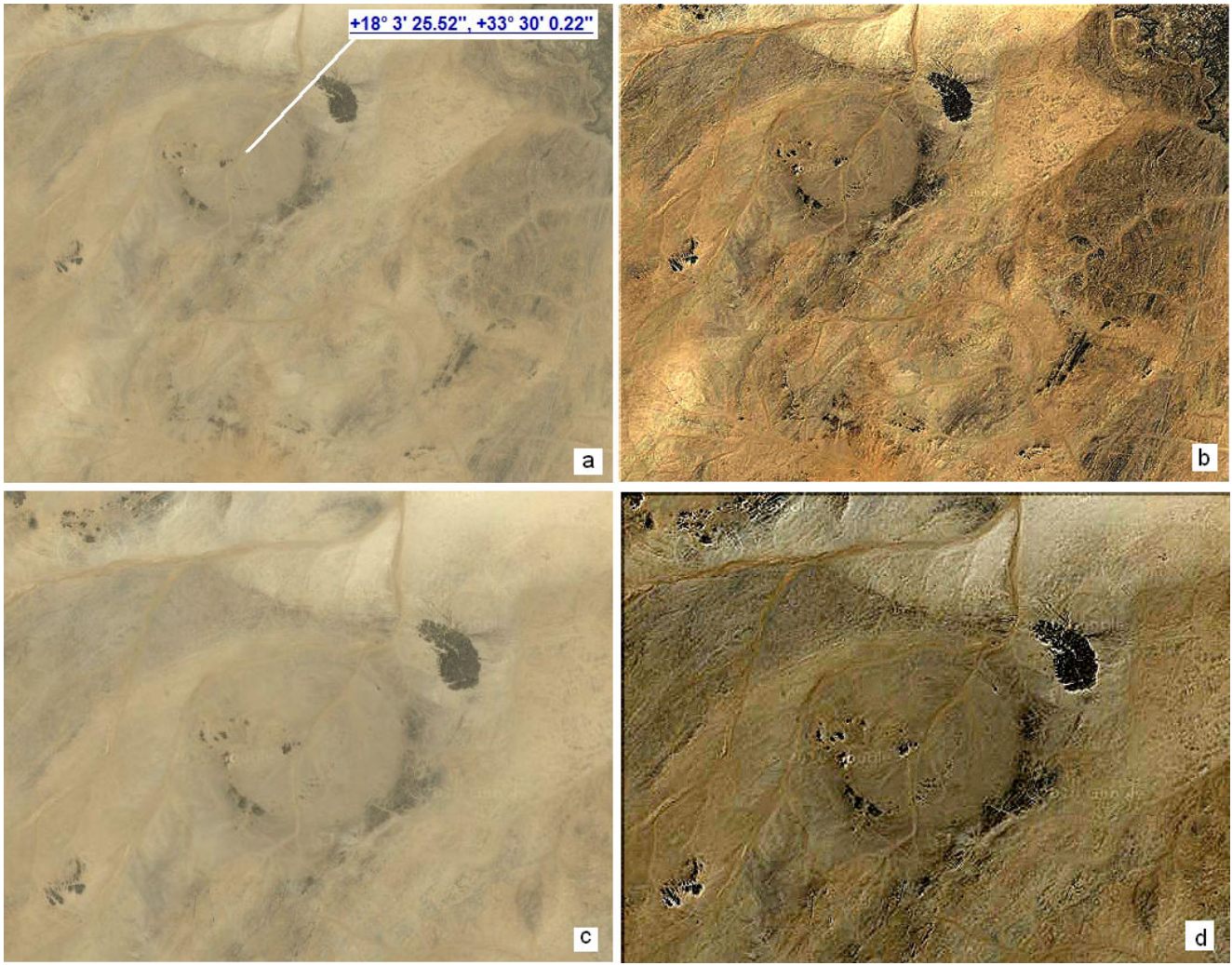

Fig.3 The image shows in detail the region on the west side of river Nile (see Fig.2). The crater-like landform, with its coordinates marked, in the upper left panel (a) has a diameter of approximately 10 km. This structure is 40 km west of Berber town. The panel (a) shows the image as obtained from Google Maps. On the right (b), the image after enhancement with AstroFracTool and Gimp. Note the embossment effect due to the use of AstroFracTool. It seems that at contact with this crater-like structure, there is another ringed structure, with a subtle bright edge, but this form could be a not genuine one. The two lower panels are showing the crater enlarged: on the left (c) the original image and on the right (d) the processed one.

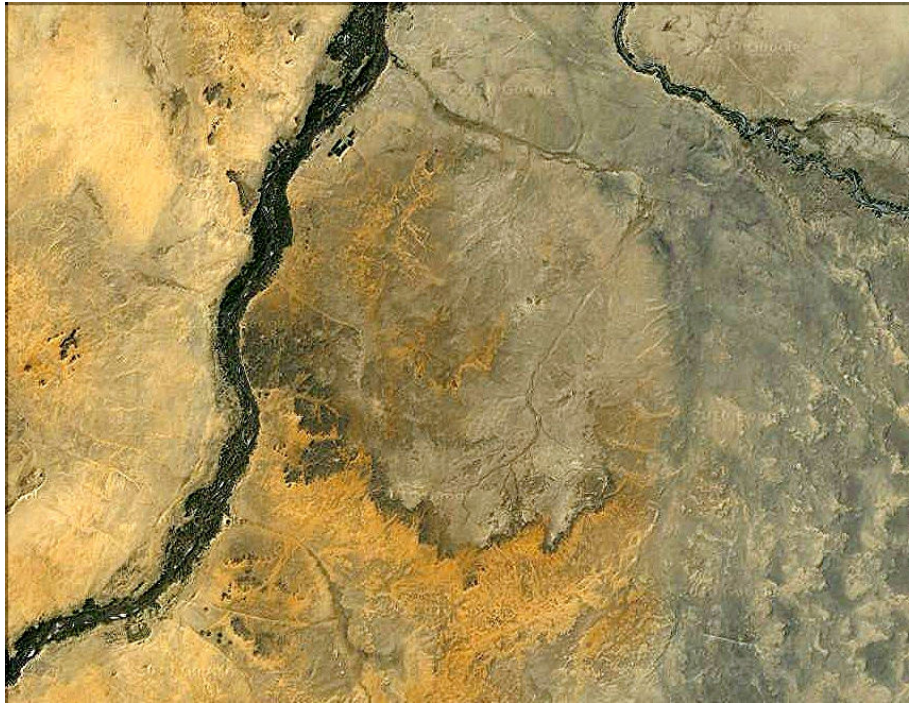

Fig.4 This is a detail of the region on the east side of Nile which is not a perfectly round (approx. 40 km diameter). This image was obtained processing a Google Map.